\documentclass[12pt]{iopart}
 \usepackage{graphicx}

\begin{document}

\title
{Anomalous viscosity of an expanding quark-gluon plasma}

\author{M Asakawa$^1$, S A Bass$^2$, and B M\"uller$^2$}

\address{$^1$ Department of Physics, Osaka University,
Toyonaka, 560-0043, Japan\\
$^2$ Department of Physics, Duke University, Durham, NC 27708, USA
}
\ead{yuki@phys.sci.osaka-u.ac.jp}
\begin{abstract}
We argue that an expanding quark-gluon plasma has an anomalous viscosity,
which arises from interactions with dynamically generated colour fields.
The anomalous viscosity dominates over the collisional
viscosity for large velocity gradients or weak coupling.
This effect may provide an explanation for the apparent
near perfect liquidity of the matter produced in nuclear collisions at
RHIC without the assumption that it is a strongly coupled state.

\end{abstract}

\section{Introduction}

Measurements of the anisotropic collective flow of hadrons emitted
in noncentral collisions of heavy nuclei at the Relativistic Heavy 
Ion Collider (RHIC) are in remarkably good agreement with the 
predictions of ideal relativistic fluid dynamics.
The comparison between data and calculations indicates that $\eta$
cannot be much larger than the postulated lower bound 
$\eta_{\rm min} = s/4\pi$ \cite{Heinz:2001xi,Teaney:2003pb,Kovtun:2004de}.
This have led to speculations
that the matter produced at RHIC is a strongly coupled quark-gluon 
plasma (sQGP). The possible microscopic structure of such a state 
is not well understood at present 
\cite{Shuryak:2004tx,Koch:2005vg}.

Here we present an alternative mechanism that may be responsible for 
a small viscosity of a weakly coupled, but expanding quark-gluon plasma.
The new mechanism is based on the theory of particle transport in 
turbulent plasmas \cite{Dupree:1966,Dupree:1968}. Such plasmas are 
characterized by strongly excited random field modes in certain regimes 
of instability, which coherently scatter the charged particles and 
thus reduce the rate of momentum transport. 
The scattering by turbulent fields in electromagnetic plasmas
is known to greatly reduce the viscosity 
\cite{Abe:1980a,Abe:1980b} of the plasma. Following Abe and Niu 
\cite{Abe:1980b}, we call the contribution from turbulent fields 
to transport coefficients ``anomalous''.

The sufficient condition for the spontaneous formation of turbulent, 
partially coherent fields is the presence of instabilities in 
the gauge field due to the presence of charged particles. This 
condition is met in electromagnetic plasmas with an anisotropic 
momentum distribution of the charged particles \cite{Weibel:1959}, 
and it is known to be satisfied in quark-gluon plasmas with an 
anisotropic momentum distribution of thermal partons 
\cite{Mrowczynski:1988dz,Mrowczynski:1993qm,Romatschke:2003ms}.

The turbulent plasma fields induce an additional,
anomalous contribution to the viscosity, which we denote as $\eta_{\rm A}$. 
This anomalous viscosity decreases with increasing strength of the 
turbulent fields. Since the amplitude of the turbulent fields grows 
with the magnitude of the momentum anisotropy, a large anisotropy 
will lead to a small value of $\eta_{\rm A}$. Because the relaxation rates 
due to different processes are additive, the total viscosity is given by 
$ \eta^{-1} = \eta_{\rm A}^{-1} + \eta_{\rm C}^{-1} $, where
$ \eta_{\rm C} $ is the collisional shear viscosity.
This implies that $\eta_{\rm A}$ dominates the total shear viscosity, 
if it is smaller than $\eta_{\rm C}$. In that limit, the anomalous mechanism
exhibits a stable equilibrium in which the viscosity regulates itself: 
The anisotropy grows with $\eta$, but an increased anisotropy tends to 
suppress $\eta_{\rm A}$ and thus $\eta \approx \eta_{\rm A}$.

\section{Anomalous viscosity}

Here we give a heuristic derivation of $\eta_{\rm A}$.
For more systematic derivations,
see references \cite{Asakawa:2006tc,Asakawa:2006jn}.
According to classical transport theory, the shear viscosity is given 
by,
\begin{equation}
\eta \approx \frac{1}{3} n \bar{p}\lambda_{\rm f} , 
\label{eq:eta-class}
\end{equation}
where $n$ denotes the particle density, $\bar{p}$ is the thermal 
momentum, and $\lambda_{\rm f}$ is the mean free path. For a weakly coupled 
quark-gluon plasma, $n \approx 5T^3$ and $\bar{p}\approx 3T$. The mean free 
path depends on the mechanism under consideration. $\eta_{\rm C}$ is
obtained by expressing the mean free path 
in terms of the transport cross section
$ \lambda_{\rm f}^{\rm (C)} = (n\,\sigma_{\rm tr})^{-1}$.
Using the perturbative QCD expression for the transport cross section
in a quark-gluon plasma yields the result, which
agrees parametrically with the result
for the collisional shear viscosity in leading logarithmic 
approximation \cite{Arnold:2000dr}.
The anomalous viscosity is determined by the same relation 
(\ref{eq:eta-class}) for $\eta$, but the mean free path is now 
obtained by counting the number of colour field domains a thermal 
parton has to traverse in order to ``forget'' its original direction 
of motion. If we denote the field strength generically by ${\cal B}^a$ 
($a$ denotes the colour index), a single coherent domain of size 
$r_{\rm m}$ causes a momentum deflection of the order of 
$\Delta p \sim gQ^a{\cal B}^ar_{\rm m}$, where $Q^a$ is the colour charge
of the parton. If different field domains are uncorrelated, the mean 
free path due to the action of the turbulent fields is given by
\begin{equation}
\lambda_{\rm f}^{\rm (A)} = r_{\rm m}\langle(\bar{p}/\Delta p)^2\rangle
\sim \frac{\bar{p}^2}{g^2Q^2\langle{\cal B}^2\rangle r_{\rm m}} .
\label{eq:lam-A}
\end{equation}
The anomalous shear viscosity thus takes the form:
\begin{equation}
\eta_{\rm A} 
\sim \frac{n\,\bar{p}^3}{3g^2Q^2\langle{\cal B}^2\rangle r_{\rm m}}
\sim \frac{9sT^3}{4g^2Q^2\langle{\cal B}^2\rangle r_{\rm m}}.
\label{eq:eta-A-app}
\end{equation}

The argument now comes down to an estimate of the average field
intensity $\langle{\cal B}^2 \rangle$ and size $r_{\rm m}$ of a domain. 
We first note that the size is given by the characteristic wave 
length of the unstable field modes. Near thermal equilibrium, the
parameter describing the influence of hard thermal partons on the
soft colour field modes is the colour-electric screening mass 
$m_{\rm D}\sim gT$. Introducing a dimensionless parameter $\xi$ for 
the magnitude of the momentum space anisotropy \cite{Romatschke:2003ms},
the wave vector domain of unstable modes is $k^2 \leq \xi m_{\rm D}^2$.
Thus $r_{\rm m}\sim \xi^{-1/2}(gT)^{-1}$. The 
exponential growth of the unstable soft field modes is saturated,
when the nonlinearities in the Yang-Mills equation become of the 
same order as the gradient term: $g|A|\sim k$, which implies 
that the field energy in the unstable mode is of the order of
$g^2\langle{\cal B}^2\rangle \sim k^4 \sim \xi^2 m_{\rm D}^4$. The 
denominator in (\ref{eq:eta-A-app}) thus has the characteristic
size, at saturation:
\begin{equation}
g^2Q^2\langle{\cal B}^2\rangle r_{\rm m} \sim \xi^{3/2} m_{\rm D}^3
\sim \xi^{3/2}(gT)^3 .
\label{eq:gB2r}
\end{equation}
If our analysis, which requires confirmation by numerical simulations,
is correct, it gives the following relation for the anomalous  
viscosity:
\begin{equation}
\eta_{\rm A} \sim \frac{s}{g^3\xi^{3/2}} \, .
\label{eq:eta-A-app2}
\end{equation}
We conclude that $\eta_{\rm A}$ will be smaller than
$\eta_{\rm C}$, if the coupling constant $g$ is sufficiently 
small and the anisotropy parameter $\xi$ is sufficiently large.

Reference \cite{Romatschke:2003ms} uses 
the following parametrization of the anisotropic momentum distribution:
\begin{equation}
f({\mathbf p}) = f_0\left(\sqrt{p^2+\xi({\mathbf p}\cdot\hat{n})^2}\right) 
\approx f_0(p) - \frac{\xi({\mathbf p}\cdot\hat{n})^2}{2E_pT}
f_0(1\pm f_0) \, ,
\end{equation}
where $ f_0 $ is the thermal equilibrium distribution.
For the special case of boost-invariant longitudinal flow,
we obtain a relation between $\eta$
and $\xi$, which takes the form (for a massless parton gas):
\begin{equation}
\xi = 10 \frac{\eta}{s}\, \frac{|\nabla u|}{T} \, ,
\quad |\nabla u| 
\equiv \left[\frac{3}{2}(\nabla u)_{ij}(\nabla u)_{ji}\right]^{1/2}
= \frac{1}{\tau} \, .
\label{eq:grad-u}
\end{equation}
Combining (\ref{eq:eta-A-app2}) and (\ref{eq:grad-u}) and
setting $\eta = \eta_A$, we obtain:
\begin{equation}
\frac{\eta_{\rm A}}{s} 
= \bar{c}_0 \left(\frac{T}{g^2|\nabla u|}\right)^{3/5} \, .
\label{eq:eta-A}
\end{equation}

The viscous contribution
to the stress tensor is proportional to $\eta |\nabla u|$.
For the collisional viscosity, this implies 
that the stress tensor grows linearly with $|\nabla u|$. The anomalous
shear viscosity (\ref{eq:eta-A}), on the other hand, is a decreasing 
function of the velocity gradient; its contribution to the stress 
tensor grows like $|\nabla u|^{2/5}$ for our scaling assumptions.
The unusual dependence of $\eta_{\rm A}$ on $|\nabla u|$ certainly 
justifies the term ``anomalous viscosity''.

\begin{figure}[thb]
\begin{center}
\vspace*{0.1cm}
\includegraphics
[width=0.38 \linewidth]
{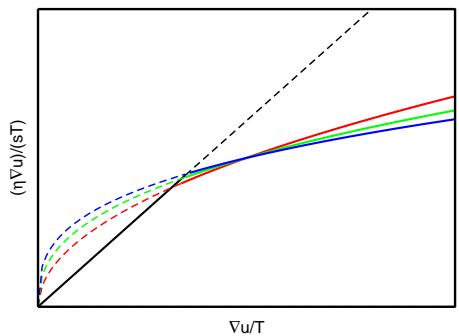}
\end{center}
\caption{
Schematic representation of the dependence of the collisional
and anomalous viscous stress on the velocity gradient.
The collisional viscous stress is shown by the linearly rising
line; the anomalous viscous stress is shown by the curved 
lines for three scaling exponents of the turbulent colour field energy 
($n=1.5,2,2.5$). The solid lines indicate the dominant source of viscous 
stress in different regions of the scaled velocity gradient $|\nabla u|/T$.
}
\label{fig1}
\vspace*{-0.7cm}
\end{figure}

The different dependence of the collisional and the anomalous viscous
stress tensors on the velocity gradient is shown schematically in 
figure~\ref{fig1}. For very small gradients the linear dependence of 
the collisional viscous stress tensor dominates, but for larger 
velocity gradients the lower power associated with the anomalous 
shear viscosity will assert its dominance. The precise location of 
the cross-over between the two domains depends on the value of the
numerical constant $\bar{c}_0$, but we can deduce from (\ref{eq:eta-A})
that the cross-over point shifts to lower values of $|\nabla u|$ with
decreasing coupling constant $g$. We also show in the figure the 
effect of choosing a different power $n$ in the scaling law,
$ g^2 \langle{\cal B}^2 \rangle = b_0 g^4 T^4 \xi^n $
for the energy density of the turbulent colour fields.
Finally, we note that the Kovtun-Son-Starinets (KSS) bound $\eta  
\geq s/4\pi$ \cite{Kovtun:2004de}
does not apply to the anomalous viscosity.

We have discussed the anomalous
viscosity in an anisotropically expanding quark-gluon plasma.
By reducing the shear viscosity of a weakly coupled, but expanding
quark-gluon plasma, this mechanism could possibly explain the observations 
of the RHIC experiments without the assumption of a strongly coupled plasma 
state. 

\ack
This work was supported in part by  
grants from the U.~S.~Department of Energy (DE-FG02-05ER41367), 
the National Science Foundation (NSF-INT-03-35392), and the 
Japanese Ministry of Education (grant-in-aid 17540255).

\end{document}